\documentclass{article}
\usepackage{Unified_TS-ASR,times}
\usepackage{amsmath}
\usepackage{amssymb}
\usepackage{hyperref}
\hypersetup{
    hypertexnames=false,
    pdftitle={Unified Target-Speaker ASR with Text and Enrollment Speech Cues},
    pdfauthor={Yuxiang Mei, Yuchen Yan, Dongxing Xu, Jiaen Liang, and Yanhua Long}
}
\usepackage{url}
\usepackage{booktabs}
\usepackage{tabularx}
\usepackage{graphicx}
\usepackage{float}
\usepackage{placeins}
\usepackage{enumitem}
\usepackage{multirow}
\usepackage{colortbl}
\usepackage{makecell}


\title{Unified \mbox{Target-Speaker} ASR with Text and \mbox{Enrollment} Speech Cues}
\author{Yuxiang Mei\textsuperscript{1,$\dagger$}, Yuchen Yan\textsuperscript{2}, Dongxing Xu\textsuperscript{3}, Jiaen Liang\textsuperscript{3}, Yanhua Long\textsuperscript{1,*} \\
Shanghai Normal University\textsuperscript{1} \quad Baidu\textsuperscript{2} \quad Unisound\textsuperscript{3} \\
\texttt{m153517@icloud.com, yanyuchen01@baidu.com} \\
\texttt{\{xudongxing,liangjiaen\}@unisound.com, yanhua@shnu.edu.cn}}

\newcommand{\cueT}{\mathrm{text}}
\newcommand{\cueE}{\mathrm{enroll}}

\iclrfinalcopy
\begin{document}
\raggedbottom
\maketitle
\lhead{Preprint}
\begingroup
\renewcommand{\thefootnote}{\fnsymbol{footnote}}
\footnotetext[1]{Corresponding author.}
\footnotetext[2]{Work done during an internship at Baidu.}
\endgroup

\begin{abstract}
Target-speaker automatic speech recognition (TS-ASR) aims to recognize speech from a designated speaker while suppressing interfering speakers in multi-talker environments. Conventional TS-ASR systems typically rely on an enrollment utterance to specify the target speaker, requiring additional speech from the same speaker at inference time. Text-guided approaches provide an alternative by exploiting known lexical content, such as a wake word, to locate the target speaker directly from the observed speech. However, these two forms of target guidance are typically studied independently, despite providing complementary information about the target speaker. In this work, we propose a Unified Dual-Cue TS-ASR framework that accommodates text cues, enrollment speech, or their combination within a single model. The text cue interacts with the mixture representation to extract target-speaker information conditioned on known lexical content, while an independent enrollment utterance provides complementary speaker information from a different utterance of the same speaker. Cross-attention cue-conditioning modules are embedded inside the shared Conformer blocks to condition ASR on the designated speaker and suppress interfering speech. The shared conditioning interface supports inference with text cues, enrollment speech, or their combination. During dual-cue training, both modalities are supplied and negative-cue sampling provides cue-validity supervision. Experiments on 30,000 two-speaker evaluation mixtures cover five recording and domain conditions and four oracle text-cue lengths. With five-character text cues, the proposed concatenated dual-cue method achieves an overall character error rate (CER) of 8.80\%, compared with 17.32\% and 29.06\% under text-only and enrollment-speech-only inference, respectively. It also outperforms parallel dual-cue fusion, which obtains 9.49\% CER, and yields lower dual-cue CER across all five evaluation subsets. These results demonstrate the advantage of jointly exploiting complementary lexical and speaker information for target-speaker ASR.
\end{abstract}

\section{Introduction}
\label{sec:introduction}

Target-speaker automatic speech recognition (TS-ASR) aims to transcribe only a designated speaker from a multi-talker mixture. Unlike multi-speaker ASR, which produces transcripts for every source, TS-ASR uses a target cue to resolve whose speech should be transcribed. A common solution uses an enrollment utterance as an acoustic reference~\citep{delcroix2019speakerbeam,moriya2022streaming,zhang2023conformer}. Although effective, this formulation requires an additional recording from the target speaker and often compresses the enrollment signal into a fixed-dimensional representation.

Known lexical content provides another form of target specification. A keyword or short phrase spoken in the mixture can anchor the target to a linguistic event without requiring a prerecorded voice~\citep{shi2024keyword,hao2023typing,li2026detect}. Text and enrollment speech provide different evidence: the former connects the target to observed content, whereas the latter supplies an independent description of speaker identity. Existing systems, however, generally treat these cues as separate interfaces and cannot exploit their complementarity within one ASR model.

Neither cue is universally reliable. A short text cue may be shared by multiple speakers, obscured by overlap, or insufficient for stable target tracking. Enrollment speech may differ from the mixture in channel, device, noise, or speaking style. When both cues are available, lexical and acoustic evidence can compensate for these ambiguities; when only one is available, the ASR interface should remain well defined. This motivates a unified formulation that accommodates text cues, enrollment speech, and their combination while preserving the information specific to each modality.

Flexible target specification also requires controlled evaluation. Text-cue length changes the amount of disclosed lexical evidence, and recording-device variation affects the reliability of the enrollment reference. Evaluating different cue configurations on the same set of mixtures enables a direct comparison of their contributions. Full-transcript ASR performance must also be measured because detecting or reproducing a keyword alone does not guarantee that the surrounding speech has been attributed to the correct speaker.

We propose a Unified Dual-Cue TS-ASR framework with a shared conditioning interface for lexical and acoustic cues. On a benchmark of 30,000 two-speaker evaluation mixtures, the concatenated unified system obtains an overall full-transcript CER of 8.80\% with five-character text cues and enrollment speech, compared with 17.32\% and 29.06\% when text or enrollment is supplied alone. During training, both modalities are supplied and negative-cue sampling independently varies their validity, encouraging the model to reject invalid cues and exploit the remaining valid information. At inference, the same checkpoint operates with text cues, enrollment speech, or both. Our main contributions are:
\begin{itemize}[leftmargin=1.5em]
    \item We propose a unified architecture with a shared conditioning interface for lexical and acoustic target cues, together with negative-cue sampling for cue-validity supervision.
    \item We introduce sequential, embedding-free enrollment conditioning that preserves frame-level acoustic information for target-speaker ASR.
    \item We construct and release the Unified-TS-ASR-Eval benchmark for evaluating text cues of flexible lengths, enrollment-speech cues, and their combination across mixtures from different recording devices. The code\footnote{\url{https://github.com/YuCeong-May/Unified-TS-ASR}}, models\footnote{\url{https://huggingface.co/YuCeong-May/Unified-TS-ASR}}, and test data\footnote{\url{https://huggingface.co/datasets/YuCeong-May/Unified-TS-ASR-Eval}} are available through anonymous repositories.
\end{itemize}
\begin{figure}[t]
    \centering
    \includegraphics[width=0.95\linewidth]{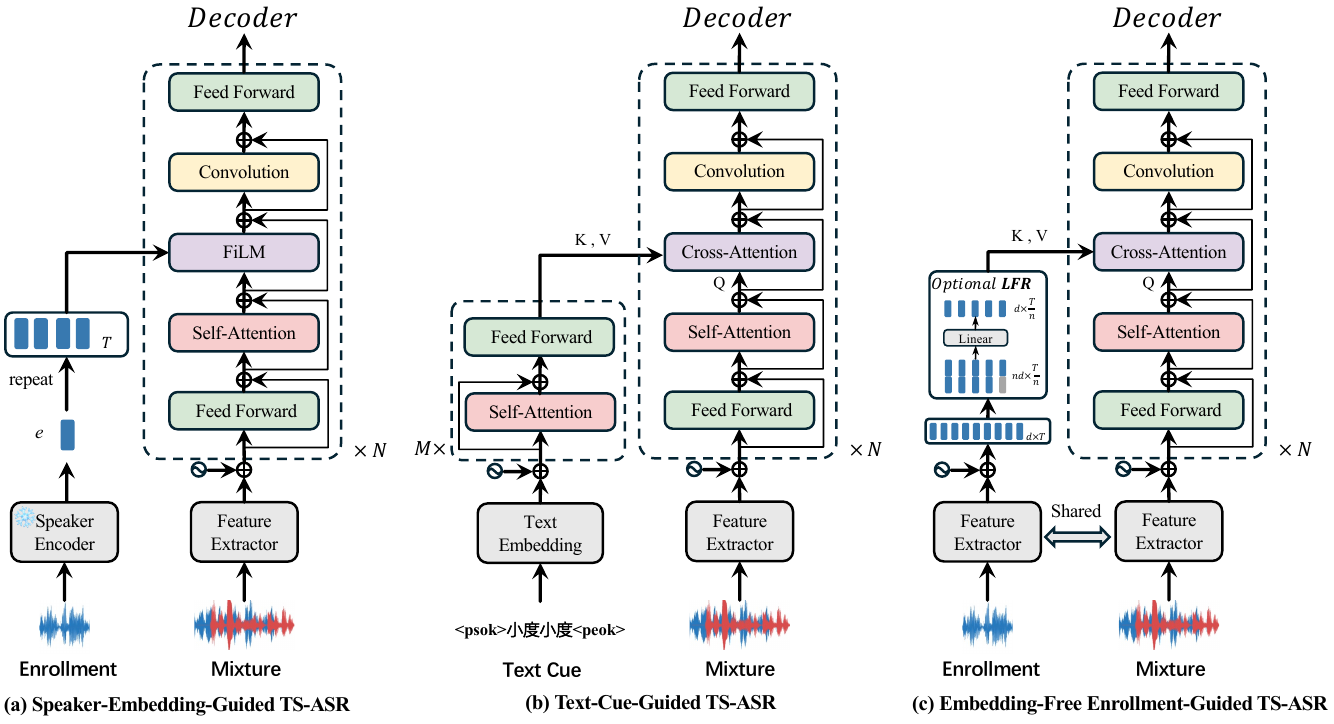}
    \caption{Overview of three single-cue TS-ASR interfaces. (a) Speaker-Embedding-Guided TS-ASR summarizes enrollment speech into a fixed-dimensional speaker representation and conditions the acoustic encoder through FiLM. (b) Text-Cue-Guided TS-ASR encodes a keyword or short phrase into token-level features for cross-attention conditioning inside each Conformer block. (c) Embedding-Free Enrollment-Guided TS-ASR retains sequential enrollment features for cross-attention conditioning inside each Conformer block.}
    \label{fig:single-cue}
\end{figure}
\section{Methodology}
\label{sec:method}

Figure~\ref{fig:single-cue} summarizes the three single-cue TS-ASR systems implemented in this work. Speaker-Embedding-Guided TS-ASR and Text-Cue-Guided TS-ASR represent two established target-conditioning paradigms, whereas Embedding-Free Enrollment-Guided TS-ASR is our proposed single-cue alternative. Building on the embedding-free design, we further introduce a Unified Dual-Cue TS-ASR framework that supports text, enrollment speech, or both through a common conditioning interface. The cue-conditioning module is embedded inside each Conformer block. In the following descriptions, $\mathbf U\in\mathbb R^{T\times d}$ denotes the intermediate mixture hidden sequence entering this internal module in a generic block, and its conditioned output is returned to the remaining sublayers of the same block.

\paragraph{\textbf{(a) Speaker-Embedding-Guided TS-ASR.}}
A pretrained speaker encoder compresses the enrollment utterance into a fixed-dimensional vector $\mathbf e=f_{\mathrm{spk}}(r)$. As shown in Figure~\ref{fig:single-cue}(a), $\mathbf e$ is used to estimate a feature-wise linear modulation (FiLM)~\citep{perez2018film} for each mixture frame:
\begin{equation}
    \mathbf H^{\mathrm{emb}}_t
    =w(\mathbf e)\odot\mathbf U_t+b(\mathbf e),
    \qquad t=1,\ldots,T.
    \label{eq:embedding-guided}
\end{equation}
where $w(\cdot)$ and $b(\cdot)$ are two learned linear projection layers, and $\odot$ denotes element-wise multiplication. Collecting the frame-level outputs gives $\mathbf H^{\mathrm{emb}}\in\mathbb R^{T\times d}$, the speaker-embedding-conditioned mixture sequence returned to the remaining sublayers of the current Conformer block. This FiLM conditioning is repeated inside the encoder blocks, is global within each block, and discards the temporal structure of the enrollment utterance.

\paragraph{\textbf{(b) Text-Cue-Guided TS-ASR.}}
As illustrated in Figure~\ref{fig:single-cue}(b), token embeddings and a Transformer text encoder~\citep{vaswani2017attention} map the known keyword or phrase $\mathbf c$ into the lexical sequence $\mathbf C=f_{\mathrm{txt}}(\mathbf c)$. The mixture representation interacts with the text-cue sequence through cross-attention (CA), followed by a residual connection and layer normalization (LN). Specifically, $\mathbf U$ supplies the queries, while $\mathbf C$ supplies the keys and values:
\begin{equation}
    \mathbf H^{\mathrm{text}}
    =\operatorname{LN}\!\left(
    \mathbf U+\operatorname{CA}(\mathbf U,\mathbf C,\mathbf C)
    \right).
    \label{eq:text-guided}
\end{equation}
Here, $\mathbf H^{\mathrm{text}}\in\mathbb R^{T\times d}$ is the text-conditioned mixture sequence returned to the remaining sublayers of the current Conformer block. This interaction is repeated inside the encoder blocks and identifies the speaker associated with the supplied lexical cue without requiring its temporal location.

\paragraph{\textbf{(c) Proposed Embedding-Free Enrollment-Guided TS-ASR.}}
Compressing enrollment speech into a single vector may discard frame-level acoustic information that is useful for speaker discrimination. In Figure~\ref{fig:single-cue}(c), our embedding-free design processes the enrollment utterance with the acoustic frontend fully shared with the mixture branch. Optional low-frame-rate (LFR) processing~\citep{pundak16_interspeech} stacks adjacent enrollment frames and projects them back to width $d$, producing the sequential enrollment representation $\mathbf R=f_{\mathrm{enroll}}(r)$. The mixture and enrollment sequences interact directly through
\begin{equation}
    \mathbf H^{\mathrm{enroll}}
    =\operatorname{LN}\!\left(
    \mathbf U+\operatorname{CA}(\mathbf U,\mathbf R,\mathbf R)
    \right),
    \label{eq:embedding-free-guided}
\end{equation}
Here, $\mathbf H^{\mathrm{enroll}}\in\mathbb R^{T\times d}$ is the sequential-enrollment-conditioned mixture sequence returned to the remaining sublayers of the current Conformer block. This interaction is repeated inside the encoder blocks and preserves frame-level enrollment information without requiring an external speaker encoder or speaker embedding.

Based on design (c), the proposed unified framework incorporates text representations into the same sequence-conditioning interface, enabling a single ASR model to operate with text cues, enrollment speech, or their combination. The task formulation and unified conditioning mechanism are described in the following subsections.

\subsection{Unified Dual-Cue TS-ASR Task Definition}
\label{sec:problem-formulation}

Let a single-channel mixture $x(\tau)$ contain speech from $K$ speakers together with additive background noise:
\begin{equation}
    x(\tau)=\sum_{k=1}^{K}s_k(\tau)+n(\tau),
    \label{eq:mixture}
\end{equation}
where $\tau$ denotes continuous time, $s_k(\tau)$ denotes the contribution of speaker $k$ at time $\tau$, and $n(\tau)$ denotes additive noise. Given this mixture, the goal is to transcribe only the designated speaker and produce the corresponding target transcript $\mathbf y^{\star}$. We consider two complementary forms of target specification. The text cue $\mathbf c$ is a keyword or short phrase spoken by the target in the mixture, but its temporal location is not provided. The enrollment utterance $r$ is a separate recording from the same speaker and provides an acoustic reference for speaker identity.

Depending on the available input, the ASR model operates with the text cue, the enrollment utterance, or both. We represent these three modes using the availability indicator $\mathbf a=(a_{\cueT},a_{\cueE})$, where $a_{\cueT}=1$ if the text cue is supplied and $0$ otherwise, while $a_{\cueE}=1$ if enrollment speech is supplied and $0$ otherwise:
\begin{equation}
    \mathcal A=\{(1,0),(0,1),(1,1)\}.
    \label{eq:modes}
\end{equation}
Here, $\mathcal A$ is the set of supported availability states: $(1,0)$ denotes text-only input, $(0,1)$ denotes enrollment-speech-only input, and $(1,1)$ denotes joint text--enrollment input. The no-cue case $(0,0)$ is excluded because the mixture alone does not specify which speaker should be transcribed. The TS-ASR problem is therefore written as
\begin{equation}
    \widehat{\mathbf y}=\arg\max_{\mathbf y}
    p_{\theta}(\mathbf y\mid x,\mathbf c,r,\mathbf a),
    \label{eq:recognition}
\end{equation}
where $\widehat{\mathbf y}$ is the predicted transcript, $\mathbf y$ ranges over candidate token sequences, and $\theta$ denotes all trainable model parameters. The indicator $\mathbf a$ determines whether $\mathbf c$ and $r$ participate in the computation; unavailable modalities are omitted. Cue availability is distinct from cue validity, which describes whether a supplied cue correctly identifies the target.

An acoustic frontend $f_{\mathrm{aud}}$ extracts frame-level features, which initialize the mixture sequence processed by $L$ cue-conditioned Conformer blocks~\citep{gulati2020conformer}:
\begin{equation}
    \mathbf U^{(0)}=f_{\mathrm{aud}}(x),\qquad
    \mathbf U^{(\ell)}
    =f_{\mathrm{conf}}^{(\ell)}
    \!\left(\mathbf U^{(\ell-1)};\mathbf C,\mathbf R,\mathbf a\right),
    \quad \ell=1,\ldots,L.
    \label{eq:conformer}
\end{equation}
Here, $\mathbf U^{(0)}\in\mathbb R^{T\times d}$ is the initial mixture sequence, and $f_{\mathrm{conf}}^{(\ell)}$ denotes the $\ell$-th Conformer block, which contains the cue-conditioning cross-attention module described below. Within that module, the current mixture hidden sequence supplies the queries, and the available cue sequences supply the keys and values according to $\mathbf a$. The final encoder representation under availability state $\mathbf a$ is denoted by $\mathbf U_{\mathbf a}^{(L)}$.

\subsection{Unified Dual-Cue TS-ASR Framework}
\label{sec:fusion}

\begin{figure}[t]
    \centering
    \includegraphics[width=1.0\linewidth]{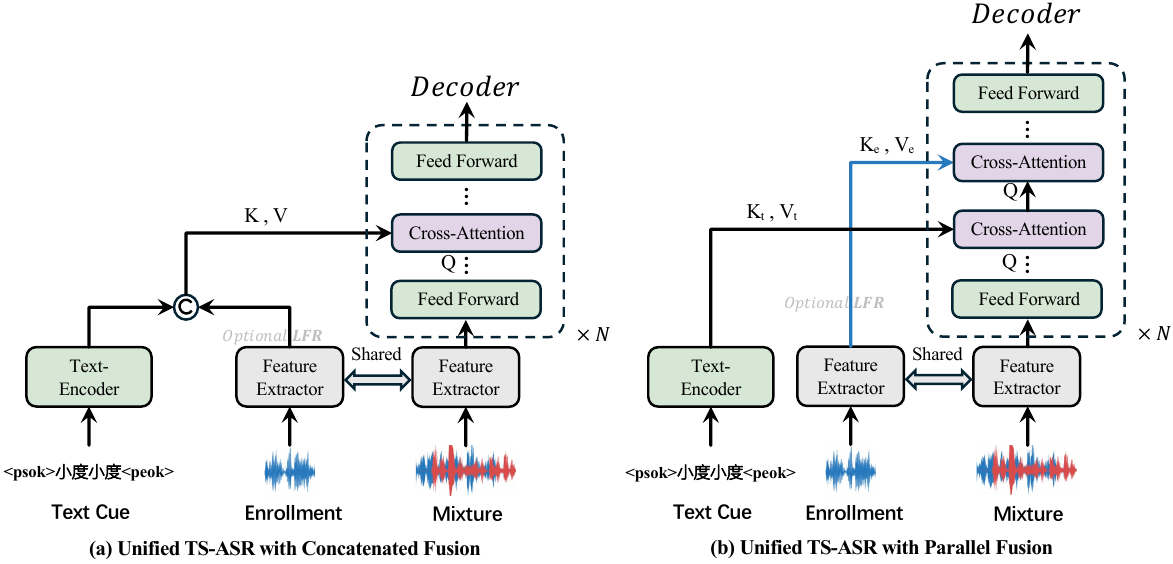}
    \caption{Two Unified Dual-Cue TS-ASR fusion designs embedded inside the Conformer blocks. (a) Concatenated fusion combines the text-cue and enrollment-speech sequences as the keys and values of one cross-attention module in each block. (b) Parallel fusion retains separate cross-attention stages for the text cue and enrollment speech in each block. In both designs, the current mixture hidden sequence supplies the queries, and the final cue-conditioned Conformer representation feeds a general ASR decoder.}
    \label{fig:dual-cue}
\end{figure}

Given a mixture waveform $x$, the acoustic frontend produces the initial mixture representation $\mathbf U^{(0)}$, while the text-cue and enrollment-speech encoders produce $\mathbf C$ and $\mathbf R$, respectively. Figure~\ref{fig:dual-cue} shows two fusion designs whose cue-conditioning modules are embedded inside each Conformer block. In both designs, the mixture and enrollment-speech branches share the acoustic feature extractor, and optional LFR shortens the enrollment-speech sequence.

\paragraph{Concatenated fusion.}
The text-cue and enrollment-speech representations are concatenated along the sequence dimension and jointly used as the keys and values of one cross-attention module in each encoder block. The mixture representation provides the queries. This design places both modalities in a common conditioning sequence and uses the same attention projections for text-cue and enrollment-speech positions.

\paragraph{Parallel fusion.}
The text-cue and enrollment-speech representations remain separate and are processed by two cue-specific cross-attention modules in each encoder block. The first module conditions the mixture stream on the text-cue representation, and the second further conditions it on the enrollment-speech representation. The two modalities therefore retain separate attention parameters before the resulting representation is passed to the remaining encoder layers and the downstream decoder.

\subsection{Unified Model Training}
\label{sec:training}

Figure~\ref{fig:dual-cue} shows the cue-conditioning modules embedded inside the repeated Conformer blocks, followed by a general ASR decoder. In this work, we instantiate the decoder with a connectionist temporal classification (CTC) output layer~\citep{graves2006ctc}. Its vocabulary contains the transcript tokens, the CTC blank, two boundary labels for a valid text cue, and two validity labels for invalid text and enrollment cues. The auxiliary labels provide no acoustic timestamps and are removed before CER computation.

Dual-cue training always supplies both modalities, giving $\mathbf a_{\mathrm{train}}=(1,1)$. We define a negative cue as a supplied cue that does not identify the designated target: a negative text cue is lexically mismatched with the target utterance, while a negative enrollment cue comes from a non-target speaker. Without such examples, the model could learn to trust every supplied cue unconditionally and fail when a cue is incorrect or conflicts with the other modality. We therefore sample text and enrollment validity independently as $\mathbf v=(v_{\cueT},v_{\cueE})\in\{0,1\}^2$. When at least one cue is valid, the target transcript is retained and any invalid modality receives cue-validity supervision; when both cues are invalid, the target contains only rejection supervision. This encourages the model to assess each cue and rely on the valid evidence. Appendix~\ref{app:cue-validity-targets} gives the exact structured CTC targets for all four combinations.

Let $\mathbf z_{\mathbf v}^{\star}$ denote the structured CTC target determined by cue-validity state $\mathbf v$. Using $\mathbf U_{\mathbf a}^{(L)}$ for the final encoder representation under availability state $\mathbf a$, the training objective is
\begin{equation}
    \mathcal{L}
    =
    \mathbb{E}_{\mathbf v\sim p(\mathbf v)}
    \left[
        \mathcal{L}_{\mathrm{CTC}}
        \left(
            \mathbf z_{\mathbf v}^{\star},
            \mathbf U_{\mathbf a_{\mathrm{train}}}^{(L)}
        \right)
    \right],
    \label{eq:training-objective}
\end{equation}
where $p(\mathbf v)$ is the independently sampled cue-validity distribution. Training varies cue validity within the dual-cue training configuration rather than changing the training availability state. The same unified checkpoint supports text-only, enrollment-speech-only, and dual-cue inference modes by omitting the unavailable cue branch.

\section{Experiments}
\label{sec:experiments}
\subsection{Datasets}
\label{sec:data}

Unified-TS-ASR-Eval is built from the official AISHELL-1~\citep{bu2017aishell1} and AISHELL-2~\citep{du2018aishell2} releases. We retain the source recordings and metadata while constructing the two-speaker mixtures, target assignments, text cues, and enrollment associations. The training splits provide approximately 1.129 million utterances from 2,331 speakers, and the development splits provide approximately 23,000 recordings from 45 speakers for validation; the AISHELL-2 development data contain parallel recordings across three devices. Further source-corpus details are provided in Appendix~\ref{app:dataset-details}.

\paragraph{Mixture construction.}
Training mixtures are generated online using the target--interferer, text-cue, and enrollment construction described below, with additional invalid cues for negative-cue sampling. Each evaluation mixture combines a designated target utterance with an utterance from a different speaker, without additional background noise. The evaluation set contains 30,000 mixtures divided evenly among five subsets, whose source domains and recording conditions are summarized in Table~\ref{tab:test-subsets}.

\begin{table*}[t]
    \caption{Overview of the Unified-TS-ASR-Eval subsets.}
    \label{tab:test-subsets}
    \centering
    \small
    \setlength{\tabcolsep}{7pt}
    \renewcommand{\arraystretch}{1.12}

    \begin{tabular}{@{}cccccc@{}}
        \toprule
        \textbf{Tag} &
        \textbf{Source Dataset} &
        \textbf{Split} &
        \textbf{Device} &
        \textbf{Pairing} &
        \textbf{Num.} \\
        \midrule

        A1
        & AISHELL-1
        & Test
        & Mic
        & Within-domain
        & 6,000 \\

        iOS
        & AISHELL-2
        & Test
        & iOS
        & Within-domain
        & 6,000 \\

        And.
        & AISHELL-2
        & Test
        & Android
        & Within-domain
        & 6,000 \\

        Mic
        & AISHELL-2
        & Test
        & Mic
        & Within-domain
        & 6,000 \\

        Mix
        & AISHELL-1 + AISHELL-2
        & Test
        & Mixed
        & Cross-domain
        & 6,000 \\

        \bottomrule
    \end{tabular}
\end{table*}

The three AISHELL-2 subsets reuse the same target--interferer pairs and mixing conditions across iOS, Android, and microphone recordings, allowing device variation to be examined under matched source content. The source signals are mixed using
\begin{equation}
    \eta \sim \operatorname{TruncatedNormal}(0,4^2,-10,10)\ \mathrm{dB},\qquad
    q \sim \operatorname{Uniform}(0.3,1.0),
    \label{eq:sir-overlap}
\end{equation}
where $\eta$ is the signal-to-interference ratio (SIR) in decibels and $q$ is the overlap ratio. In $\operatorname{TruncatedNormal}(0,4^2,-10,10)$, the underlying Gaussian has location $0$~dB and variance $4^2$~dB$^2$ (equivalently, a standard deviation of $4$~dB), and is truncated to the interval $[-10,10]$~dB. The distribution $\operatorname{Uniform}(0.3,1.0)$ samples $q$ uniformly between $0.3$ and $1.0$. Relative source offsets are randomized.

\paragraph{Text-cue construction.}
\label{sec:cue-protocol}

For each target utterance, a contiguous oracle span of two to five characters (\texttt{cue2}--\texttt{cue5}) is selected from the transcript without providing its temporal location. Shorter utterances retain the available span, and all cue-length conditions reuse the same mixture, target--interferer pair, and mixing condition.

\paragraph{Enrollment construction.}
\label{sec:enrollment-protocol}

Enrollment speech is selected from a different utterance of the target speaker as an independent acoustic reference. The unified and embedding-free enrollment models use the same enrollment source recordings.

\subsection{Experimental Setup}
\label{sec:experimental-setup}

\paragraph{Cue configuration and decoding.}
All systems use greedy CTC decoding without an external language model, using the checkpoint with the highest development-set accuracy. The unified models are trained with both modalities supplied and with cue validity independently randomized through negative-cue sampling, enabling the model to reject an invalid cue and rely on the remaining valid information. At inference, the same checkpoint is evaluated in text-only, enrollment-speech-only, and dual-cue modes; unavailable cues are omitted from the conditioning computation.

For the speaker-embedding system, we use the publicly released CAM++ speaker encoder~\citep{wang2023campp}, pretrained on a large-scale Mandarin corpus containing 200,000 labeled speakers.

Detailed model and optimization configurations are provided in Appendix~\ref{app:configuration}.

For the text-cue systems, N0, N10, N30, and N50 denote negative-text sampling rates of 0\%, 10\%, 30\%, and 50\%, respectively. Negative-text examples supply a lexical cue that is invalid with respect to the designated target utterance and supervise the corresponding cue-validity label.

We report two corpus-level transcription metrics: full-transcript CER and outside-cue CER. Full-transcript CER is computed over the complete target transcript, including the cue characters, whereas outside-cue CER scores only the remaining non-cue characters as detailed in Appendix~\ref{app:scoring}. Before either metric is computed, the reference and hypothesis are normalized using the same rules for Chinese and English text, with punctuation and auxiliary cue-boundary or cue-validity labels removed.

\subsection{Main Results}
\label{sec:main-results}

\begin{table*}[t]
    \caption{Full-transcript CER (\%) on five test subsets for the Enrollment-Speech-Guided TS-ASR (E1--E3), Text-Cue-Guided TS-ASR (T1--T4), and Unified Dual-Cue TS-ASR (U1--U6) systems.
    A1 denotes AISHELL-1; iOS, And., and Mic denote the AISHELL-2 channels;
    Mix denotes the cross-domain test set. The Input column uses Text, Enroll, and Text + Enroll for the three cue configurations.
    Text-N$\rho$ denotes a text-cue model trained with a negative-text sampling rate of $\rho\%$, where $\rho\in\{0,10,30,50\}$.
    For Text and Text + Enroll inputs, the text-cue length is fixed at 5 characters.
    Bold indicates the best result for each test subset and overall.}
    \label{tab:main-cer}
    \centering
    \small
    \setlength{\tabcolsep}{6pt}
    \renewcommand{\arraystretch}{1.05}

    \begin{tabular}{@{}clcrrrrrr@{}}
        \toprule
        \textbf{ID} & \textbf{System} & \textbf{Input} &
        \textbf{A1} & \textbf{iOS} & \textbf{And.} &
        \textbf{Mic} & \textbf{Mix} & \textbf{All} \\
        \midrule

        \rowcolor{gray!20}
        \multicolumn{9}{c}{\textbf{Enrollment-Speech-Guided TS-ASR}} \\
        \addlinespace[1pt]

        E1 & Speaker-Embedding
            & Enroll & \textbf{4.95} & 16.98 & 21.69 & 18.62 & 19.75 & 15.50 \\
        E2 & Embedding-free
            & Enroll & 5.17 & 18.30 & 23.16 & 19.89 & 20.36 & 16.41 \\
        E3 & Embedding-free (LFR)
            & Enroll & 5.51 & 17.32 & 22.44 & 17.66 & 21.58 & 16.04 \\

        \addlinespace[2pt]
        \rowcolor{gray!20}
        \multicolumn{9}{c}{\textbf{Text-Cue-Guided TS-ASR}} \\
        \addlinespace[1pt]

        T1 & Text-N0
            & Text & 26.59 & 28.60 & 29.56 & 28.23 & 25.61 & 27.57 \\
        T2 & Text-N10
            & Text & 17.34 & 15.97 & 17.17 & 15.84 & 15.17 & 16.36 \\
        T3 & Text-N30
            & Text & 15.16 & 15.88 & 17.32 & 15.31 & 13.79 & 15.43 \\
        T4 & Text-N50
            & Text & 15.29 & 16.19 & 18.08 & 15.71 & 14.17 & 15.80 \\

        \addlinespace[2pt]
        \rowcolor{gray!20}
        \multicolumn{9}{c}{\textbf{Unified Dual-Cue TS-ASR}} \\
        \addlinespace[1pt]

        U1 & Parallel
            & Enroll & 9.76 & 24.04 & 28.24 & 23.14 & 39.68 & 24.00 \\
        U2 & Parallel
            & Text & 8.75 & 15.46 & 18.05 & 14.25 & 10.96 & 13.04 \\
        U3 & Parallel
            & Text + Enroll & 5.58 & 11.07 & 13.05 & 10.33 & 9.31 & 9.49 \\
        \addlinespace[1pt]

        U4 & Concat
            & Enroll & 15.84 & 26.64 & 33.60 & 27.08 & 46.31 & 29.06 \\
        U5 & Concat
            & Text & 12.72 & 19.77 & 23.31 & 18.26 & 15.02 & 17.32 \\
        U6 & Concat
            & Text + Enroll & 5.12 & \textbf{9.94} & \textbf{12.06} &
                    \textbf{9.31} & \textbf{9.27} & \textbf{8.80} \\

        \bottomrule
    \end{tabular}
\end{table*}

\paragraph{Single-cue specialist systems.}
Table~\ref{tab:main-cer} shows that negative-text sampling substantially improves the text-cue systems, reducing overall CER from 27.57\% for Text-N0 (T1) to 15.43\% for Text-N30 (T3). The full-frame and LFR embedding-free enrollment systems obtain 16.41\% and 16.04\% CER, respectively, approaching the 15.50\% achieved by the speaker-embedding system.

\paragraph{Unified systems.}
Both architectures benefit substantially from combining the two cues: Parallel improves from 13.04\% and 24.00\% under text-only and enrollment-speech-only inference to 9.49\% with both cues, while Concat improves from 17.32\% and 29.06\% to 8.80\%. Concat achieves the best dual-cue CER and outperforms Parallel on all five evaluation subsets, whereas Parallel is more robust under either single-cue input.

Parallel's single-cue advantage is unlikely to arise from competition between modalities because an unavailable cue is omitted. Instead, its modality-specific cross-attention modules reduce the input-distribution change from dual-cue training: single-cue inference activates the branch specialized for the available cue, whereas Concat applies one shared module to conditioning sequences with different compositions and lengths. These results characterize inference modes of the same unified checkpoint rather than separately trained single-cue systems.

\subsection{Text-Cue-Length and Enrollment Analysis}
\label{sec:ablation-analysis}

\begin{table}[t]
    \caption{Overall full-transcript and outside-cue CER (\%) versus requested text-cue length on paired mixtures. Text + Enroll keeps enrollment fixed, and shorter transcripts use the available span. Scoring follows Appendix~\ref{app:scoring}; system IDs follow Table~\ref{tab:main-cer}. Bold indicates the best result at each length.}
    \label{tab:cue-length}
    \centering
    \small
    \setlength{\tabcolsep}{3pt}
    \renewcommand{\arraystretch}{1.05}

    \begin{tabular}{@{}clcrrrrrrrr@{}}
        \toprule
        \multirow{2}{*}{\textbf{ID}} &
        \multirow{2}{*}{\textbf{System}} &
        \multirow{2}{*}{\textbf{Input}} &
        \multicolumn{4}{c}{\textbf{Full-transcript CER}}
          & \multicolumn{4}{c}{\textbf{Outside-cue CER}} \\
        \cmidrule(lr){4-7}\cmidrule(lr){8-11}
        & & &
        \textbf{Len.=2} & \textbf{Len.=3} & \textbf{Len.=4} & \textbf{Len.=5} &
        \textbf{Len.=2} & \textbf{Len.=3} & \textbf{Len.=4} & \textbf{Len.=5} \\
        \midrule

        \rowcolor{gray!20}
        \multicolumn{11}{c}{\textbf{Text-Cue-Guided TS-ASR}} \\
        \addlinespace[1pt]

        T1 & Text-N0  & Text & 44.57 & 37.69 & 32.35 & 27.57 & 52.24 & 49.55 & 48.15 & 45.81 \\
        T2 & Text-N10 & Text & 28.58 & 23.26 & 19.44 & 16.36 & 33.77 & 30.85 & 29.30 & 27.85 \\
        T3 & Text-N30 & Text & 28.33 & 22.35 & 18.49 & 15.43 & 33.29 & 29.50 & 27.67 & 26.13 \\
        T4 & Text-N50 & Text & 30.10 & 23.24 & 19.01 & 15.80 & 35.12 & 30.57 & 28.43 & 26.72 \\

        \addlinespace[2pt]
        \rowcolor{gray!20}
        \multicolumn{11}{c}{\textbf{Unified Dual-Cue TS-ASR}} \\
        \addlinespace[1pt]

        U2 & Parallel & Text & 27.16 & 20.25 & 15.98 & 13.04 & 31.11 & 25.95 & 23.28 & 21.01 \\
        U3 & Parallel & Text + Enroll & 15.69 & 13.37 & 11.33 & 9.49 & 18.18 & 17.40 & 16.66 & 15.52 \\
        \addlinespace[1pt]
        U5 & Concat & Text & 29.59 & 24.23 & 20.43 & 17.32 & 34.14 & 31.26 & 29.79 & 27.95 \\
        U6 & Concat & Text + Enroll & \textbf{14.59} & \textbf{12.55} & \textbf{10.61} & \textbf{8.80}
                         & \textbf{17.09} & \textbf{16.46} & \textbf{15.72} & \textbf{14.67} \\

        \bottomrule
    \end{tabular}
\end{table}

\paragraph{Effect of text-cue length.}
As shown in Table~\ref{tab:cue-length}, both full-transcript and outside-cue CER decrease consistently as the oracle text cue becomes longer. For the concatenated unified system U6, increasing the text-cue length from 2 to 5 characters reduces full-transcript CER from 14.59\% to 8.80\% and outside-cue CER from 17.09\% to 14.67\%. The outside-cue improvement shows that the benefit is not limited to reproducing the disclosed cue characters; the additional lexical context also improves ASR for the remaining target speech. The same monotonic trend holds for all text-only systems (T1--T4, U2, and U5) and the Parallel system U3, while U6 gives the lowest CER under both scoring conventions at every evaluated length. Because the cue is selected from the scored transcript, these gains may still combine easier target localization with improved speaker tracking. Additional text-conditioned training variants are reported in Appendix~\ref{app:additional-results}.

\paragraph{Embedding-free enrollment variants.}
We implement the proposed embedding-free enrollment system in two variants. E2 retains the original frame rate of the sequential enrollment representation and achieves 16.41\% overall CER, whereas E3 applies LFR before cross-attention and achieves 16.04\%. The LFR variant adds only 0.192 million parameters (0.65\%) relative to E2, and both variants are smaller than the 35.076-million-parameter speaker-embedding system E1. A complete comparison of ASR performance and model size is provided in Appendix~\ref{app:model-parameters}.

\subsection{Visualization Analysis}
\label{sec:visualization-analysis}

\begin{figure}[t]
    \centering
    \includegraphics[width=0.8\linewidth]{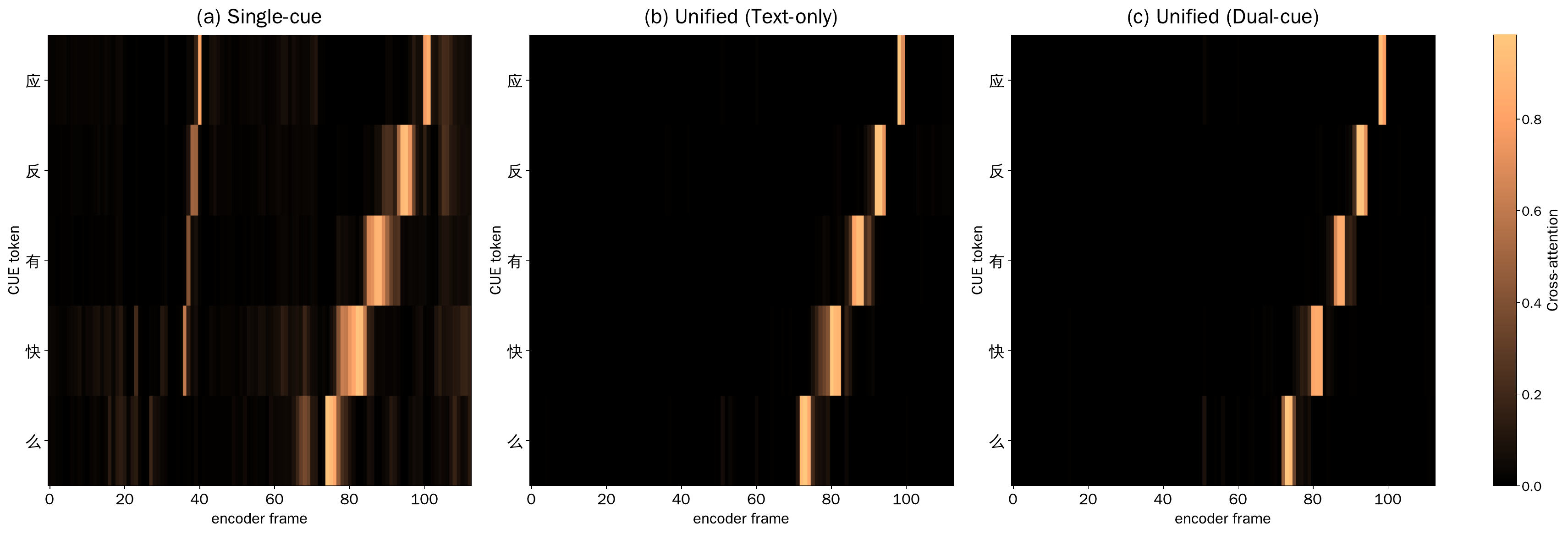}
    \caption{Text-side cross-attention for the same example with (a) the single-cue text model, (b) text-only inference using the unified model, and (c) joint text--enrollment inference using the same unified model. Columns show encoded mixture frames, rows show text-cue tokens, and brighter regions denote larger attention weights.}
    \label{fig:text-attn}
\end{figure}

Figure~\ref{fig:text-attn} visualizes text-side cross-attention for the same evaluation example under three conditioning settings. The single-cue text model exhibits relatively diffuse interactions over the mixture sequence, whereas text-only inference with the unified model yields more concentrated token--frame interactions. With both text and enrollment cues, the text-side attention becomes further concentrated over a smaller portion of the mixture sequence. This change is consistent with the two cues providing complementary conditioning signals within the unified model.

These patterns provide clear evidence of learned cue–mixture interactions rather than explicit temporal supervision, and should therefore not be interpreted as supervised keyword localization.

\section{Related Work}
\label{sec:related}

\paragraph{Enrollment-conditioned TS-ASR.}
Enrollment-conditioned systems specify the target through a reference utterance. Cascaded approaches first extract the target waveform, whereas direct systems inject enrollment-derived speaker representations into the acoustic model~\citep{wang2019voicefilter,delcroix2019speakerbeam}. Subsequent work extends this formulation to generative extraction, streaming transducers, Conformer encoders, and foundation-model-based ASR~\citep{ma2025generative,moriya2022streaming,zhang2023conformer,ma2024prompt,guo2025sqwhisper,polok2025whisper,polok2026sedicow,kim2026llmasr}. These methods establish enrollment speech as a practical target interface but generally rely on explicit speaker representations.

\paragraph{Embedding-free and sequential enrollment conditioning.}
Speaker information has been incorporated into self-supervised ASR and target-speech pre-training through speaker-aware representations~\citep{huang2023adapting,zhang2023tshubert}. In target-speech extraction and personalized enhancement, sequential enrollment features preserve temporal acoustic information and interact directly with the mixture~\citep{yang2024contextual,zeng2025useftse,huang2025sefpnet}. Our embedding-free branch follows this sequential-conditioning direction, shares the acoustic frontend between mixture and enrollment speech, and applies frame-level enrollment conditioning directly to target-speaker transcription.

\paragraph{Text- and keyword-guided TS-ASR.}
Text-guided target-speech extraction uses linguistic descriptions or partial transcripts to identify the desired source, while keyword-guided TS-ASR uses an in-mixture keyword as a target-selection anchor~\citep{hao2023typing,shi2024keyword,li2026detect}. Unlike contextual biasing, which primarily modifies lexical decoding probabilities, these approaches associate a known lexical event with the target speaker~\citep{pundak2018deep,shakeel2026calm}. Existing work, however, generally studies lexical and enrollment-speech cues as separate target interfaces; our framework unifies them within one model and supports either cue or their combination.

\section{Conclusion}
\label{sec:conclusion}

We proposed a unified target-speaker ASR framework that integrates text cues and enrollment speech through a shared sequence-conditioning interface. By representing both cue types as conditioning sequences, the framework provides a common inference interface for text-guided, enrollment-guided, and joint text--enrollment ASR. Dual-cue training supplies both modalities and uses negative-cue sampling to supervise whether each supplied cue validly identifies the target. Results on 30,000 two-speaker evaluation mixtures demonstrate that the two cues provide complementary information for target-speaker ASR. With five-character text cues, the concatenated dual-cue model achieves an overall CER of 8.80\%, outperforming the parallel fusion design and substantially improving over its text-only and enrollment-only inference modes. We further show that longer text cues consistently improve ASR performance and that low-frame-rate enrollment encoding introduces only a small parameter overhead.

Overall, the results indicate that lexical and acoustic speaker information can be effectively combined within a common conditioning framework for target-speaker ASR. Future work will investigate explicit single-cue training, non-oracle text cues, and broader multi-speaker scenarios.

\section*{The Use of Large Language Models}
Generative AI tools were used to assist with literature retrieval and discovery, research ideation and implementation, drafting and language editing, and code refinement. All AI-assisted outputs were critically reviewed and verified by the authors, who take full responsibility for the methodology, experiments, analysis, and final content of this work.

\bibliography{Unified_TS-ASR}

@inproceedings{wang2019voicefilter,
  title = {{VoiceFilter}: Targeted Voice Separation by Speaker-Conditioned Spectrogram Masking},
  author = {Wang, Quan and Muckenhirn, Hannah and Wilson, Kevin and Sridhar, Prashant and Wu, Zelin and Hershey, John and Saurous, Rif A. and Weiss, Ron J. and Jia, Ye and Moreno, Ignacio Lopez},
  booktitle = {Proc. Interspeech},
  year = {2019},
  pages = {2728--2732},
  doi = {10.21437/Interspeech.2019-1101}
}

@inproceedings{delcroix2019speakerbeam,
  title = {End-to-End {SpeakerBeam} for Single Channel Target Speech Recognition},
  author = {Delcroix, Marc and Watanabe, Shinji and Ochiai, Tsubasa and Kinoshita, Keisuke and Karita, Shigeki and Ogawa, Atsunori and Nakatani, Tomohiro},
  booktitle = {Proc. Interspeech},
  year = {2019},
  pages = {451--455},
  doi = {10.21437/Interspeech.2019-1856}
}

@inproceedings{moriya2022streaming,
  title = {Streaming Target-Speaker {ASR} with Neural Transducer},
  author = {Moriya, Takafumi and Sato, Hiroshi and Ochiai, Tsubasa and Delcroix, Marc and Shinozaki, Takahiro},
  booktitle = {Proc. Interspeech},
  year = {2022},
  pages = {2673--2677}
}

@inproceedings{huang2023adapting,
  title = {Adapting Self-Supervised Models to Multi-Talker Speech Recognition Using Speaker Embeddings},
  author = {Huang, Zili and Raj, Desh and Garc{\'i}a, Paola and Khudanpur, Sanjeev},
  booktitle = {Proc. IEEE International Conference on Acoustics, Speech and Signal Processing (ICASSP)},
  year = {2023},
  pages = {1--5}
}

@inproceedings{yang2024contextual,
  title = {Target Speaker Extraction by Directly Exploiting Contextual Information in the Time-Frequency Domain},
  author = {Yang, Xue and Bao, Changchun and Zhou, Jing and Chen, Xianhong},
  booktitle = {Proc. IEEE International Conference on Acoustics, Speech and Signal Processing (ICASSP)},
  year = {2024}
}

@article{zeng2025useftse,
  title = {{USEF-TSE}: Universal Speaker Embedding Free Target Speaker Extraction},
  author = {Zeng, Bang and Li, Ming},
  journal = {IEEE Transactions on Audio, Speech, and Language Processing},
  volume = {33},
  pages = {2110--2124},
  year = {2025},
  doi = {10.1109/TASLPRO.2025.3572756}
}

@inproceedings{huang2025sefpnet,
  title = {{SEF-PNet}: Speaker Encoder-Free Personalized Speech Enhancement with Local and Global Contexts Aggregation},
  author = {Huang, Ziling and Guan, Haixin and Wei, Haoran and Long, Yanhua},
  booktitle = {Proc. IEEE International Conference on Acoustics, Speech and Signal Processing (ICASSP)},
  year = {2025}
}

@article{hao2023typing,
  title = {Typing to Listen at the Cocktail Party: Text-Guided Target Speaker Extraction},
  author = {Hao, Xiang and Wu, Jibin and Yu, Jianwei and Xu, Chenglin and Tan, Kay Chen},
  journal = {IEEE Transactions on Cognitive and Developmental Systems},
  volume = {18},
  number = {2},
  pages = {361--372},
  year = {2026},
  doi = {10.1109/TCDS.2025.3598687}
}

@inproceedings{li2026detect,
  title = {Detect, Attend and Extract: Keyword Guided Target Speaker Extraction},
  author = {Li, Haoyu and Xi, Yu and Jiang, Yidi and Wang, Shuai and Knill, Kate and Gales, Mark and Li, Haizhou and Yu, Kai},
  booktitle = {Proc. Thirty-Fifth International Joint Conference on Artificial Intelligence (IJCAI)},
  year = {2026},
  pages = {5784--5792}
}

@inproceedings{shakeel2026calm,
  title = {{CALM}: Joint Contextual Acoustic-Linguistic Modeling for Personalization of Multi-Speaker {ASR}},
  author = {Shakeel, Muhammad and Fukumoto, Yosuke and Maeda, Chikara and Lin, Chyi-Jiunn and Watanabe, Shinji},
  booktitle = {Proc. IEEE International Conference on Acoustics, Speech and Signal Processing (ICASSP)},
  year = {2026},
  doi = {10.1109/ICASSP55912.2026.11463102}
}

@inproceedings{gulati2020conformer,
  title = {{Conformer}: Convolution-Augmented Transformer for Speech Recognition},
  author = {Gulati, Anmol and Qin, James and Chiu, Chung-Cheng and Parmar, Niki and Zhang, Yu and Yu, Jiahui and Han, Wei and Wang, Shibo and Zhang, Zhengdong and Wu, Yonghui and Pang, Ruoming},
  booktitle = {Proc. Interspeech},
  year = {2020},
  pages = {5036--5040},
  doi = {10.21437/Interspeech.2020-3015}
}

@article{ma2025generative,
  title = {Enhancing Intelligibility for Generative Target Speech Extraction via Joint Optimization with Target Speaker {ASR}},
  author = {Ma, Hao and Chen, Rujin and Zhang, Xiao-Lei and Liu, Ju and Li, Xuelong},
  journal = {IEEE Signal Processing Letters},
  volume = {32},
  pages = {2309--2313},
  year = {2025},
  doi = {10.1109/LSP.2025.3573951}
}

@inproceedings{zhang2023conformer,
  title = {Conformer-Based Target-Speaker Automatic Speech Recognition for Single-Channel Audio},
  author = {Zhang, Yang and Puvvada, Krishna C. and Lavrukhin, Vitaly and Ginsburg, Boris},
  booktitle = {Proc. IEEE International Conference on Acoustics, Speech and Signal Processing (ICASSP)},
  year = {2023},
  pages = {1--5}
}

@inproceedings{zhang2023tshubert,
  title = {Weakly-Supervised Speech Pre-training: A Case Study on Target Speech Recognition},
  author = {Zhang, Wangyou and Qian, Yanmin},
  booktitle = {Proc. Interspeech},
  year = {2023},
  pages = {3517--3521}
}

@inproceedings{ma2024prompt,
  title = {Extending {Whisper} with Prompt Tuning to Target-Speaker {ASR}},
  author = {Ma, Hao and Peng, Zhiyuan and Shao, Mingjie and Li, Jing and Liu, Ju},
  booktitle = {Proc. IEEE International Conference on Acoustics, Speech and Signal Processing (ICASSP)},
  year = {2024},
  pages = {12516--12520}
}

@article{guo2025sqwhisper,
  title = {{SQ-Whisper}: Speaker-Querying Based {Whisper} Model for Target-Speaker {ASR}},
  author = {Guo, Pengcheng and Chang, Xuankai and Lv, Hang and Watanabe, Shinji and Xie, Lei},
  journal = {IEEE Transactions on Audio, Speech, and Language Processing},
  volume = {33},
  pages = {175--185},
  year = {2025},
  doi = {10.1109/TASLP.2024.3513835}
}

@inproceedings{polok2025whisper,
  title = {Target Speaker {ASR} with {Whisper}},
  author = {Polok, Alexander and Klement, Dominik and Wiesner, Matthew and Khudanpur, Sanjeev and {\v C}ernock{\'y}, Jan and Burget, Luk{\'a}{\v s}},
  booktitle = {Proc. IEEE International Conference on Acoustics, Speech and Signal Processing (ICASSP)},
  year = {2025},
  pages = {1--5}
}

@inproceedings{polok2026sedicow,
  title = {{SE-DiCoW}: Self-Enrolled Diarization-Conditioned {Whisper}},
  author = {Polok, Alexander and Klement, Dominik and Cornell, Samuele and Wiesner, Matthew and {\v C}ernock{\'y}, Jan and Khudanpur, Sanjeev and Burget, Luk{\'a}{\v s}},
  booktitle = {Proc. IEEE International Conference on Acoustics, Speech and Signal Processing (ICASSP)},
  year = {2026}
}

@inproceedings{kim2026llmasr,
  title = {Target-Speaker {LLM-ASR} with Speaker-Aware Speech Encoder},
  author = {Kim, Minsoo and Kim, SangHun},
  booktitle = {Proc. IEEE International Conference on Acoustics, Speech and Signal Processing (ICASSP)},
  year = {2026},
  pages = {16732--16736}
}

@article{shi2024keyword,
  title = {Keyword Guided Target Speech Recognition},
  author = {Shi, Ying and Li, Lantian and Wang, Dong and Han, Jiqing},
  journal = {IEEE Signal Processing Letters},
  volume = {31},
  pages = {1945--1949},
  year = {2024}
}

@inproceedings{graves2006ctc,
  title = {Connectionist Temporal Classification: Labelling Unsegmented Sequence Data with Recurrent Neural Networks},
  author = {Graves, Alex and Fern{\'a}ndez, Santiago and Gomez, Faustino and Schmidhuber, J{\"u}rgen},
  booktitle = {Proc. International Conference on Machine Learning},
  year = {2006},
  pages = {369--376},
  doi = {10.1145/1143844.1143891}
}

@inproceedings{bu2017aishell1,
  title = {{AISHELL-1}: An Open-Source Mandarin Speech Corpus and a Speech Recognition Baseline},
  author = {Bu, Hui and Du, Jiayu and Na, Xingyu and Wu, Bengu and Zheng, Hao},
  booktitle = {Proc. 20th Conference of the Oriental Chapter of the International Coordinating Committee on Speech Databases and Speech I/O Systems and Assessment (O-COCOSDA)},
  year = {2017},
  pages = {1--5}
}

@misc{du2018aishell2,
  title = {{AISHELL-2}: Transforming Mandarin {ASR} Research into Industrial Scale},
  author = {Du, Jiayu and Na, Xingyu and Liu, Xuechen and Bu, Hui},
  howpublished = {arXiv preprint arXiv:1808.10583},
  year = {2018},
  eprint = {1808.10583},
  archivePrefix = {arXiv},
  primaryClass = {cs.CL},
  url = {https://arxiv.org/abs/1808.10583}
}

@inproceedings{pundak2018deep,
  title = {Deep Context: End-to-End Contextual Speech Recognition},
  author = {Pundak, Golan and Sainath, Tara N. and Prabhavalkar, Rohit and Kannan, Anjuli and Zhao, Ding},
  booktitle = {Proc. IEEE Spoken Language Technology Workshop (SLT)},
  year = {2018}
}

@inproceedings{vaswani2017attention,
  title = {Attention Is All You Need},
  author = {Vaswani, Ashish and Shazeer, Noam and Parmar, Niki and Uszkoreit, Jakob and Jones, Llion and Gomez, Aidan N. and Kaiser, Lukasz and Polosukhin, Illia},
  booktitle = {Proc. Advances in Neural Information Processing Systems},
  year = {2017}
}

@inproceedings{perez2018film,
  title = {{FiLM}: Visual Reasoning with a General Conditioning Layer},
  author = {Perez, Ethan and Strub, Florian and de Vries, Harm and Dumoulin, Vincent and Courville, Aaron},
  booktitle = {Proc. AAAI Conference on Artificial Intelligence},
  year = {2018}
}

@inproceedings{wang2023campp,
  title = {{CAM++}: A Fast and Efficient Network for Speaker Verification Using Context-Aware Masking},
  author = {Wang, Hui and Zheng, Siqi and Chen, Yafeng and Cheng, Luyao and Chen, Qian},
  booktitle = {Proc. Interspeech},
  year = {2023},
  pages = {5301--5305},
  doi = {10.21437/Interspeech.2023-1513}
}

@inproceedings{pundak16_interspeech,
  title = {{Lower Frame Rate Neural Network Acoustic Models}},
  author = {Golan Pundak and Tara N. Sainath},
  year = {2016},
  booktitle = {Proc. Interspeech 2016},
  pages = {22--26},
  doi = {10.21437/Interspeech.2016-275},
  issn = {2958-1796}
}
\bibliographystyle{Unified_TS-ASR}

\FloatBarrier
\appendix
\section{Appendix}

\subsection{Dataset Details}
\label{app:dataset-details}

Unified-TS-ASR-Eval draws its source utterances from AISHELL-1 and AISHELL-2. AISHELL-1 provides microphone recordings, whereas AISHELL-2 uses iOS recordings for training and parallel iOS, Android, and microphone recordings for development and testing. The official training, development, and test partitions serve as the sources for model training, validation, and evaluation, respectively. Table~\ref{tab:source-data} summarizes their sizes.

\begin{table}[t]
    \caption{Source-corpus statistics. Counts are rounded to the nearest thousand.}
    \label{tab:source-data}
    \centering
    \small
    \begin{tabular}{@{}lcccc@{}}
        \toprule
        Corpus & Split & Channel & Utterances & Speakers \\
        \midrule
        \multirow{3}{*}{AISHELL-1~\citep{bu2017aishell1}}
        & Train & \multirow{3}{*}{Mic} & 120K & 340 \\
        & Dev & & 14K & 40 \\
        & Test & & 7K & 20 \\
        \midrule
        \multirow{3}{*}{AISHELL-2~\citep{du2018aishell2}}
        & Train & iOS & 1,009K & 1,991 \\
        & Dev & \multirow{2}{*}{iOS, Android, Mic} & 3K each & 5 \\
        & Test & & 5K each & 10 \\
        \bottomrule
    \end{tabular}
\end{table}

\subsection{Model and Training Configuration}
\label{app:configuration}

\paragraph{Model configuration.}
All systems use 80-dimensional log-Mel filterbank features extracted from 16-kHz audio with a 25-ms frame length and a 10-ms frame shift. The acoustic encoder contains nine Conformer blocks with a hidden dimension of 256. Each block uses four-head self-attention and a convolution module with kernel size 15. The text-cue encoder is a four-layer Transformer with a hidden dimension of 256 and four attention heads. The text encoder and CTC output layer use the same character vocabulary, augmented with the auxiliary symbols \texttt{<psok>}, \texttt{<peok>}, \texttt{<no\_kws/>}, and \texttt{<no\_tgt\_spk/>}. In the speaker-embedding system, the 192-dimensional speaker embedding is mapped by separate linear layers to 256-dimensional FiLM scale and bias vectors. For embedding-free enrollment conditioning, the enrollment and mixture branches use the same acoustic frontend with fully shared parameters. The LFR variant reduces the enrollment sequence length by a factor of three before cross-attention.

\paragraph{Training configuration.}
All systems are optimized with Adam for 100 epochs using an initial learning rate of $1\times10^{-3}$ and 10,000 warm-up steps. Gradients are clipped to a maximum norm of 5. The text-cue systems and the speaker-embedding system use dynamic token-based batches with a token budget of 50,000. The embedding-free enrollment systems, including the unified systems with Parallel and Concat fusion, use a token budget of 25,000 and accumulate gradients over two steps, giving the same effective token budget of 50,000. Each microbatch contains at most 200 samples. Source and target sequence lengths are limited to 2,000 and 210, respectively, and text-cue lengths are sampled from two to five characters. All experiments are trained on four NVIDIA RTX 4090 GPUs with 24~GB of memory per GPU using DeepSpeed ZeRO Stage~1. Unless otherwise specified, the compared systems use the same optimization and training configuration. For the unified models, the validity state $\mathbf v=(v_{\cueT},v_{\cueE})$ defined in Section~\ref{sec:training} is sampled independently as $v_{\cueT}\sim\operatorname{Bernoulli}(0.7)$ and $v_{\cueE}\sim\operatorname{Bernoulli}(0.9)$, where a value of 1 denotes a valid cue. Equivalently, the text and enrollment negative-cue probabilities are 30\% and 10\%, respectively. Negative text cues are mismatched with the target utterance, while negative enrollment utterances are sampled from non-target speakers.

\subsection{Cue-Validity Target Details}
\label{app:cue-validity-targets}

Table~\ref{tab:cue-validity-training} specifies the structured CTC targets used for the four independently sampled cue-validity states. The boundary labels enclose a valid text cue, while the two rejection labels identify an invalid text cue or enrollment cue.

\begin{table}[t]
    \caption{Structured CTC targets for the four cue-validity states.}
    \label{tab:cue-validity-training}
    \centering
    \small
    \setlength{\tabcolsep}{4pt}
    \renewcommand{\arraystretch}{1.08}
    \begin{tabular}{@{}ccc@{}}
        \toprule
        \textbf{Text cue} & \textbf{Enrollment cue} & \textbf{CTC target} \\
        \midrule
        Valid   & Valid   & Transcript with cue boundaries \\
        Valid   & Invalid & \texttt{<no\_tgt\_spk/>} + transcript with cue boundaries \\
        Invalid & Valid   & \texttt{<no\_kws/>} + target transcript \\
        Invalid & Invalid & \texttt{<no\_kws/> <no\_tgt\_spk/>} only \\
        \bottomrule
    \end{tabular}
\end{table}

\subsection{Additional Results and Diagnostics}
\label{app:additional-results}

Table~\ref{tab:output-diagnostics} reports the output-label diagnostics for the text-cue systems and the unified models.

\begin{table}[t]
    \caption{Output-label rates (\%) over 30,000 evaluation mixtures, using \texttt{best} checkpoints and a text-cue length of 5 characters whenever text is supplied. Bracket hit checks for an ordered nonempty \texttt{<psok>}--\texttt{<peok>} pair; \texttt{No\_kws} and \texttt{No\_tgt\_spk} count emitted \texttt{<no\_kws/>} and \texttt{<no\_tgt\_spk/>} validity labels, respectively.}
    \label{tab:output-diagnostics}
    \centering
    \small
    \setlength{\tabcolsep}{6pt}
    \renewcommand{\arraystretch}{1.05}

    \begin{tabular}{@{}lcccc@{}}
        \toprule
        \textbf{System} & \textbf{Input} & \textbf{Bracket hit} &
        \textbf{\texttt{No\_kws}} & \textbf{\texttt{No\_tgt\_spk}} \\
        \midrule

        \rowcolor{gray!20}
        \multicolumn{5}{c}{\textbf{Text-Cue-Guided TS-ASR}} \\
        \addlinespace[1pt]

        Text-N0  & Text & 99.54 & 0.00 & -- \\
        Text-N10 & Text & 99.64 & 0.23 & -- \\
        Text-N30 & Text & 99.38 & 0.33 & -- \\
        Text-N50 & Text & 99.44 & 0.27 & -- \\

        \addlinespace[2pt]
        \rowcolor{gray!20}
        \multicolumn{5}{c}{\textbf{Unified Dual-Cue TS-ASR}} \\
        \addlinespace[1pt]

        Parallel & Enroll & 0.00 & 99.98 & 6.03 \\
        Parallel & Text & 99.24 & 0.13 & 99.99 \\
        Parallel & Text + Enroll & 99.46 & 0.20 & 5.35 \\
        \addlinespace[1pt]
        Concat & Enroll & 0.00 & 99.54 & 6.47 \\
        Concat & Text & 99.53 & 0.15 & 98.75 \\
        Concat & Text + Enroll & 99.68 & 0.11 & 5.30 \\
        \bottomrule
    \end{tabular}
\end{table}

\paragraph{Output-label interpretation.}
During training, \texttt{<no\_kws/>} indicates that a supplied text cue is invalid, and \texttt{<no\_tgt\_spk/>} indicates that supplied enrollment speech is from a non-target speaker. Across the four text-cue systems, bracket-hit rates range from 99.38\% to 99.64\%, while \texttt{No\_kws} rates remain between 0.00\% and 0.33\%. The unified models exhibit the expected behavior when one modality is withheld at inference. With enrollment-speech-only input, Parallel and Concat have a 0.00\% bracket-hit rate and emit \texttt{No\_kws} for 99.98\% and 99.54\% of the evaluation mixtures, respectively. With text-only input, bracket-hit rates rise to 99.24\% and 99.53\%, whereas \texttt{No\_tgt\_spk} is emitted for 99.99\% and 98.75\%. Supplying both cues preserves high bracket-hit rates (99.46\% for Parallel and 99.68\% for Concat) while reducing \texttt{No\_tgt\_spk} to 5.35\% and 5.30\%. Concat therefore shows slightly more consistent label behavior than Parallel in the dual-cue condition. These rates remain output-label diagnostics: bracket hit does not verify cue correctness or timing, and validity labels can coexist with a transcript in singly invalid training cases. They are not calibrated accept/reject or false-rejection rates.

\subsection{Model Performance and Parameter Comparison}
\label{app:model-parameters}

\begin{table}[t]
    \caption{Overall full-transcript CER and total parameter counts for the single-cue and unified architectures. For each architecture family, the best All result from Table~\ref{tab:main-cer} is reported. The Input column uses Text, Enroll, and Text + Enroll for the three cue configurations.}
    \label{tab:model-parameters}
    \centering
    \small
    \setlength{\tabcolsep}{12pt}
    \renewcommand{\arraystretch}{1.08}

    \begin{tabular}{@{}lccc@{}}
        \toprule
        \textbf{System} & \textbf{Input} & \textbf{CER (\%)} & \textbf{Params (M)} \\
        \midrule

        \rowcolor{gray!20}
        \multicolumn{4}{c}{\textbf{Enrollment-Speech-Guided TS-ASR}} \\
        \addlinespace[1pt]
        Speaker-Embedding     & Enroll & 15.50 & 35.076 \\
        Embedding-free       & Enroll & 16.41 & \textbf{29.636} \\
        Embedding-free (LFR) & Enroll & 16.04 & 29.828 \\

        \addlinespace[2pt]
        \rowcolor{gray!20}
        \multicolumn{4}{c}{\textbf{Text-Cue-Guided TS-ASR}} \\
        \addlinespace[1pt]
        Text-N30         & Text & 15.43 & 37.182 \\

        \addlinespace[2pt]
        \rowcolor{gray!20}
        \multicolumn{4}{c}{\textbf{Unified Dual-Cue TS-ASR}} \\
        \addlinespace[1pt]
        Concat       & Text + Enroll & \textbf{8.80} & 37.182 \\
        Parallel     & Text + Enroll & 9.49 & 39.565 \\

        \bottomrule
    \end{tabular}
\end{table}

Table~\ref{tab:model-parameters} compares the best overall CER and parameter count for each architecture. Among the enrollment-speech-guided systems, the embedding-free model is the smallest, with 29.636 million parameters. LFR reduces the temporal resolution of the enrollment-speech sequence, yet the overall CER improves slightly from 16.41\% to 16.04\%. This observation suggests that most speaker-discriminative information may be preserved after frame-rate reduction, while the LFR projection introduces only 0.192 million additional parameters. By comparison, the speaker-embedding system requires an additional pretrained speaker encoder and increases the model size to 35.076 million parameters, but improves CER over the LFR embedding-free system by only 0.54 percentage points. The text-cue system achieves 15.43\% CER, which is comparable to the enrollment-speech-guided results, showing that lexical evidence can provide an effective alternative form of target specification without enrollment speech. Combining the two modalities produces a substantially larger improvement: Concat and Parallel achieve 8.80\% and 9.49\% CER, respectively. The consistent advantage of Text + Enroll input over either single-cue modality demonstrates that the lexical information supplied by the text cue and the speaker information supplied by enrollment speech are complementary. Concat further achieves the best overall CER with the same 37.182 million parameters as the text-cue system and 2.383 million fewer parameters than Parallel.

\subsection{CER by Text-Cue Length and Evaluation Subset}
\label{app:cer-text-length}

Table~\ref{tab:cer-text-length-detailed} gives the complete full-transcript and outside-cue CER breakdown by requested text-cue length and evaluation subset. All text-cue-length conditions use paired mixtures; only the oracle text-cue length changes. The Input column uses Text and Text + Enroll, consistent with the main-text tables. Subset abbreviations follow Table~\ref{tab:test-subsets}, and values are rounded to two decimal places.

\begin{table}[t]
    \caption{Detailed CER (\%) by requested text-cue length and evaluation subset. \emph{Full-transcript} scores the complete target transcript (including cue characters); \emph{Outside-cue} scores only the remaining non-cue characters (Appendix~\ref{app:scoring}).}
    \label{tab:cer-text-length-detailed}
    \centering
    \small
    \setlength{\tabcolsep}{3pt}
    \renewcommand{\arraystretch}{1.02}

    \begin{tabular}{@{}lccrrrrrrrr@{}}
        \toprule
        \multirow{2}{*}{\textbf{System}} &
        \multirow{2}{*}{\textbf{Input}} &
        \multirow{2}{*}{\textbf{Subset}} &
        \multicolumn{4}{c}{\textbf{Full-transcript CER}} &
        \multicolumn{4}{c}{\textbf{Outside-cue CER}} \\
        \cmidrule(lr){4-7}\cmidrule(lr){8-11}
        & & & \textbf{Len.=2} & \textbf{Len.=3} & \textbf{Len.=4} & \textbf{Len.=5}
              & \textbf{Len.=2} & \textbf{Len.=3} & \textbf{Len.=4} & \textbf{Len.=5} \\
        \midrule

        \rowcolor{gray!20}
        \multicolumn{11}{c}{\textbf{Text-Cue-Guided TS-ASR}} \\
        \addlinespace[1pt]

        Text-N0  & Text & A1   & 37.60 & 32.80 & 29.50 & 26.59 & 43.02 & 40.92 & 40.49 & 40.84 \\
        Text-N0  & Text & iOS  & 48.78 & 40.82 & 34.45 & 28.60 & 60.29 & 57.94 & 57.30 & 57.79 \\
        Text-N0  & Text & And. & 50.50 & 42.30 & 35.57 & 29.56 & 62.23 & 59.95 & 59.24 & 59.72 \\
        Text-N0  & Text & Mic  & 48.06 & 40.03 & 33.84 & 28.23 & 59.33 & 56.75 & 56.35 & 57.22 \\
        Text-N0  & Text & Mix  & 41.62 & 35.19 & 30.05 & 25.61 & 50.10 & 47.85 & 46.94 & 47.01 \\
        \addlinespace[1pt]
        Text-N10 & Text & A1   & 25.41 & 21.26 & 19.04 & 17.34 & 28.67 & 26.04 & 25.42 & 25.18 \\
        Text-N10 & Text & iOS  & 30.57 & 24.54 & 19.80 & 15.97 & 36.86 & 33.82 & 31.74 & 29.68 \\
        Text-N10 & Text & And. & 32.54 & 26.11 & 21.22 & 17.17 & 39.27 & 36.07 & 34.11 & 31.94 \\
        Text-N10 & Text & Mic  & 29.92 & 23.86 & 19.39 & 15.84 & 36.24 & 33.05 & 31.20 & 29.47 \\
        Text-N10 & Text & Mix  & 26.25 & 21.64 & 18.10 & 15.17 & 31.19 & 28.86 & 27.48 & 25.97 \\
        \addlinespace[1pt]
        Text-N30 & Text & A1   & 23.19 & 19.03 & 16.81 & 15.16 & 26.15 & 23.28 & 22.35 & 21.92 \\
        Text-N30 & Text & iOS  & 31.26 & 24.41 & 19.67 & 15.88 & 37.53 & 33.45 & 31.28 & 29.41 \\
        Text-N30 & Text & And. & 33.73 & 26.29 & 21.29 & 17.32 & 40.51 & 36.06 & 33.91 & 31.96 \\
        Text-N30 & Text & Mic  & 30.61 & 23.51 & 19.02 & 15.31 & 36.82 & 32.42 & 30.49 & 28.37 \\
        Text-N30 & Text & Mix  & 25.62 & 20.33 & 16.66 & 13.79 & 30.16 & 26.92 & 25.05 & 23.59 \\
        \addlinespace[1pt]
        Text-N50 & Text & A1   & 25.58 & 20.04 & 17.17 & 15.29 & 28.65 & 24.44 & 22.90 & 22.36 \\
        Text-N50 & Text & iOS  & 32.73 & 24.92 & 20.12 & 16.19 & 39.11 & 34.17 & 32.02 & 29.80 \\
        Text-N50 & Text & And. & 35.78 & 27.57 & 22.27 & 18.08 & 42.47 & 37.55 & 35.25 & 32.95 \\
        Text-N50 & Text & Mic  & 31.73 & 24.34 & 19.38 & 15.71 & 37.95 & 33.46 & 31.00 & 29.09 \\
        Text-N50 & Text & Mix  & 27.16 & 21.09 & 17.18 & 14.17 & 31.77 & 27.86 & 25.86 & 24.12 \\
        \addlinespace[2pt]
        \rowcolor{gray!20}
        \multicolumn{11}{c}{\textbf{Unified Dual-Cue TS-ASR}} \\
        \addlinespace[1pt]

        Parallel & Text & A1   & 17.75 & 12.69 & 10.29 & 8.75 & 19.83 & 15.44 & 13.76 & 12.75 \\
        Parallel & Text & iOS  & 31.81 & 24.27 & 18.89 & 15.46 & 37.17 & 32.27 & 29.13 & 26.81 \\
        Parallel & Text & And. & 36.54 & 27.73 & 22.24 & 18.05 & 42.63 & 36.62 & 34.03 & 31.07 \\
        Parallel & Text & Mic  & 30.27 & 22.70 & 17.74 & 14.25 & 35.44 & 30.22 & 27.54 & 24.93 \\
        Parallel & Text & Mix  & 24.25 & 17.77 & 13.72 & 10.96 & 27.70 & 22.83 & 20.12 & 18.00 \\
        \addlinespace[1pt]
        Concat & Text & A1   & 19.98 & 16.32 & 14.35 & 12.72 & 22.56 & 20.03 & 19.25 & 18.60 \\
        Concat & Text & iOS  & 34.46 & 28.29 & 23.58 & 19.77 & 40.50 & 37.84 & 36.42 & 34.50 \\
        Concat & Text & And. & 40.07 & 32.91 & 27.67 & 23.31 & 46.63 & 43.43 & 41.92 & 39.80 \\
        Concat & Text & Mic  & 32.51 & 26.50 & 21.96 & 18.26 & 38.35 & 35.61 & 34.14 & 31.95 \\
        Concat & Text & Mix  & 25.93 & 21.22 & 17.80 & 15.02 & 30.12 & 27.64 & 26.24 & 24.52 \\
        \addlinespace[1pt]
        Parallel & Text + Enroll & A1   & 7.56 & 6.84 & 6.23 & 5.58 & 8.57 & 8.44 & 8.36 & 8.21 \\
        Parallel & Text + Enroll & iOS  & 18.16 & 15.79 & 13.31 & 11.07 & 21.59 & 21.37 & 20.83 & 19.68 \\
        Parallel & Text + Enroll & And. & 21.46 & 18.43 & 15.61 & 13.05 & 25.28 & 24.81 & 24.11 & 22.67 \\
        Parallel & Text + Enroll & Mic  & 17.43 & 14.92 & 12.53 & 10.33 & 20.84 & 20.43 & 19.90 & 18.81 \\
        Parallel & Text + Enroll & Mix  & 17.46 & 13.88 & 11.37 & 9.31 & 20.04 & 17.91 & 16.65 & 15.22 \\
        \addlinespace[1pt]
        Concat & Text + Enroll & A1   & 7.17 & 6.46 & 5.81 & 5.12 & 8.18 & 8.00 & 7.84 & 7.57 \\
        Concat & Text + Enroll & iOS  & 16.60 & 14.35 & 12.00 & 9.94 & 19.92 & 19.55 & 18.88 & 18.01 \\
        Concat & Text + Enroll & And. & 20.32 & 17.40 & 14.73 & 12.06 & 24.09 & 23.56 & 22.94 & 21.50 \\
        Concat & Text + Enroll & Mic  & 15.94 & 13.73 & 11.36 & 9.31 & 19.24 & 18.98 & 18.18 & 17.39 \\
        Concat & Text + Enroll & Mix  & 16.22 & 13.56 & 11.33 & 9.27 & 18.98 & 17.71 & 16.75 & 15.37 \\

        \bottomrule
    \end{tabular}
\end{table}

\subsection{Outside-Cue CER Scoring}
\label{app:scoring}

Outside-cue CER is computed from a deterministic minimum-edit alignment between the normalized reference and hypothesis. Reference positions within the selected cue span are excluded, while substitutions and deletions are retained only for positions outside that span. Each insertion is assigned to the following reference character, or to the final character when it occurs after the reference, and is counted only when the assigned position lies outside the cue span. Errors and non-cue reference characters are aggregated at the corpus level; evaluation mixtures with no remaining reference characters are omitted. The released scorer implements the fixed alignment tie-breaking rule used for all reported outside-cue results.

\end{document}